\documentclass[twocolumn]{aastex62}
\usepackage{multirow}
\shorttitle{Mg II h,k and triplet}
\shortauthors{Zhu et al.}

\begin{document}
\title{Modeling Mg II h, k and Triplet Lines at Solar Flare Ribbons}

\author{Yingjie Zhu}
\affiliation{School of Earth and Space Sciences, Peking University, Beijing 100871, China; huitian@pku.edu.cn, yj\_zhu@pku.edu.cn}
\author{Adam F. Kowalski}
\affiliation{Department of Astrophysical and Planetray Sciences, University of Colorado Boulder, 2000 Colorado Ave, Boulder, CO 80305,USA}
\affiliation{National Solar Observatory, University of Colorado Boulder, 3665 Discovery Drive, Boulder, CO 80303, USA}
\author{Hui Tian}
\affiliation{School of Earth and Space Sciences, Peking University, Beijing 100871, China; huitian@pku.edu.cn, yj\_zhu@pku.edu.cn}
\author{Han Uitenbroek}
\affiliation{National Solar Observatory, University of Colorado Boulder, 3665 Discovery Drive, Boulder, CO 80303, USA}
\author{Mats Carlsson}
\affiliation{Rosseland Centre for Solar Physics, University of Oslo, P.O. Box 1029 Blindern, NO-0315 Oslo, Norway}
\affiliation{Institute of Theoretical Astrophysics, University of Oslo, P.O. Box 1029 Blindern, NO-0315 Oslo, Norway}
\author{Joel C. Allred}
\affiliation{NASA/Goddard Space Flight Center, Code 671, Greenbelt, MD 20771, USA}

\begin{abstract}

Observations from the \textit{Interface Region Imaging Spectrograph} (\textsl{IRIS}) often reveal significantly broadened and non-reversed profiles of the Mg II h, k and triplet lines at flare ribbons. To understand the formation of these optically thick Mg II lines, we perform plane parallel radiative hydrodynamics modeling with the RADYN code, and then recalculate the Mg II line profiles from RADYN atmosphere snapshots using the radiative transfer code RH. We find that the current RH code significantly underestimates the Mg II h \& k Stark widths. By implementing semi-classical perturbation approximation results of quadratic Stark broadening from the STARK-B database in the RH code, the Stark broadenings are found to be one order of magnitude larger than those calculated from the current RH code. However, the improved Stark widths are still too small, and another factor of 30 has to be multiplied to reproduce the significantly broadened lines and adjacent continuum seen in observations. Non-thermal electrons, magnetic fields, three-dimensional effects or electron density effect may account for this factor. Without modifying the RADYN atmosphere, we have also reproduced non-reversed Mg II h \& k profiles, which appear when the electron beam energy flux is decreasing. These profiles are formed at an electron density of $\sim 8\times10^{14}\ \mathrm{cm}^{-3}$ and a temperature of $\sim1.4\times10^4$ K, where the source function slightly deviates from the Planck function. Our investigation also demonstrates that at flare ribbons the triplet lines are formed in the upper chromosphere, close to the formation heights of the h \& k lines.
\end{abstract}

\section{Introduction}
Solar flares are one of the most energetic processes in the solar system. They convert magnetic energy in the solar atmosphere into thermal and kinetic energy, and produce significantly enhanced broad-band electromagnetic radiation. According to the widely accepted two-dimensional (2D) CSHKP model \citep{Carmichael1964,Sturrock1966,Hirayama1974,Kopp1976}, magnetic reconnection, occurring in the corona, releases an enormous amount of energy, part of which is transported to the lower atmosphere along post-flare loops through non-thermal electrons \citep[e.g.,][]{Holman2003}, thermal conduction \citep[e.g.,][]{Qiu2013} or Alfv\'en waves \citep[e.g.,][]{Fletcher2008}. Continuous heating increases the local temperature, electron density and plasma velocity in the chromosphere and transition region (TR). Hence, some strong spectral lines from the Ca II, Mg II and Si IV ions as well as the H$\alpha$ and Ly$\alpha$ lines, which are formed in the chromosphere and TR, have become important diagnostics for the response of the lower atmosphere to energy transport and deposition (e.g., \citealt{Liu2015,Tian2015,Tei2018}).

Mg II h \& k lines ($\lambda_h = 2803.53~\mbox{\AA}, \lambda_k = 2796.35~\mbox{\AA}$; wavelengths in vacuum used throughout this paper) are among the most prominent chromospheric lines both in the quiet sun and at flare ribbons. They are resonance lines formed in transitions from the $3p\ ^2\mathrm{P}_{1/2}$ to $3s\ ^2\mathrm{S}_{1/2}$ and $3p\ ^2\mathrm{P}_{3/2}$ to $3s\ ^2\mathrm{S}_{1/2}$ orbitals. Three subordinate Mg II lines ($\lambda = 2791.60~\mbox{\AA},\ 2798.75~\mbox{\AA}$ and $2798.82~\mbox{\AA}$) are called triplets and are formed in transitions from the orbitals $3d\ ^2\mathrm{D}_{3/2}$ to $3p\ ^2\mathrm{P}_{1/2}$, $3d\ ^2\mathrm{D}_{3/2}$ to $3p\ ^2\mathrm{P}_{3/2}$ and $3d\ ^2\mathrm{D}_{5/2}$ to $3p\ ^2\mathrm{P}_{3/2}$, respectively. In the following the two blended longer-wavelength subordinate lines will be referred to as the 2798~\mbox{\AA} line while the shorter-wavelength subordinate line will be referred to as the 2791~\mbox{\AA} line. A few early observations (e.g.,\citealt{Lemaire1967,Lemaire1969,Kohl1976,Doschek1977}) revealed a central reversal for both Mg II h \& k in the quiet sun. The reversed line cores (h$_3$ and k$_3$) of the h \& k doublets are generally formed in the upper chromosphere, whereas the emission peaks (h$_2$ or k$_2$) are formed in lower layers \citep{Vernazza1981}. It has been found that partial frequency redistribution (PRD) treatment is more appropriate than complete frequency redistribution (CRD) approximation for the formation of the Mg II lines in non-local thermodynamic equilibrium (NLTE) radiative transfer simulations \citep{Mihalas1974,Ayres1976,Uitenbroek1997}.

The launch of the \textit{Interface Region Imaging Spectrograph} (\textsl{IRIS}; \citealt{DePontieu2014}) brought a revolution in observations of the Mg II lines with unprecedented spatial and temporal resolution. IRIS observations have also been used to guide numerical simulations of the Mg II lines in the quiet sun and plage regions. Three-dimensional (3D) radiative magnetohydrodynamic (RMHD) simulations have confirmed the necessity to include PRD effect \citep{Leenaarts2013a} and revealed important behavior of Mg II h, k and triplets. For instance, \citet{Leenaarts2013a} found that the 3D radiative transfer process affects Mg II h \& k line core formation. \citet{Leenaarts2013b} suggested that both the h \& k lines could be used to diagnose temperature and velocity in the upper chromosphere. Besides h \& k lines, \citet{Pereira2015a} investigated the subordinate lines and found that they are usually absorption lines formed in the lower chromosphere. However, in case of heating in the lower chromosphere, they become emission features and the core-to-wing ratio is well correlated to the temperature increase.

The Mg II lines observed at flare ribbons show many different characteristics compared to those observed in the quiet sun. The lines at flare ribbons are usually much more enhanced and broadened, revealing a single emission peak without central reversal (e.g., \citealt{Kerr2015,Tian2015}). A red wing enhancement and a typical red shift of $\sim$ 10 km s$^{-1}$ are also common for the Mg II line profiles observed at flare ribbons (e.g., \citealt{Li2015,Tian2018}). At the fronts of propagating flare ribbons, central-reversed h \& k lines that are accompanied by a dramatically enhanced red wing are also occasionally observed (e.g., \citealt{Liu2015,Panos2018}). The subordinate lines become emission lines at ribbons (e.g., \citealt{Tian2018}), possibly indicating a steep temperature increase in the lower atmosphere.

Previous 1D radiative hydrodynamic (RHD) simulations of flares with non-thermal electron heating cannot reproduce the observed Mg II spectra with self-consistent atmospheres. Through a data-constrained RHD simulation with the 1D RADYN \citep{Carlsson1992,Carlsson1997,Allred2005,Allred2015} and RH codes \citep{Uitenbroek2001,Pereira2015b}, \citet{Rubio2016} found that the Mg II h \& k lines are always reversed and much narrower than in observations. A following parameter study \citep{Rubio2017} has managed to reproduce single-peak Mg II h \& k lines by manually increasing temperature or electron density at the line core formation height to maintain a local thermodynamic equilibrium (LTE) condition. \cite{Rubio2017} also concluded that the widely broadened line profiles might be reproduced by combining unresolved up- and down-flows with speeds up to $\sim$ 250 km s$^{-1}$, though the exact shapes of the line profiles are still different from those in observations. In another simulation where the flare ribbon is heated by Alfv\'en waves, \citet{Kerr2016} found that the calculated Mg II k line profile evolves from being central-reversed into single-peaked. However, the profiles are strongly asymmetric and again lack far wing emission.

Recently, \citet{Kowalski2017a} revisited the X1.1-class flare on 2014 March 29. The near-ultraviolet (NUV) continuum enhancement and asymmetric Fe II line profiles in their 1D RHD simulation with RADYN and RH show a good consistency with the IRIS observation. Moreover, they found that the RH code may underestimate the electron pressure broadening of spectral lines. In this paper, we perform 1D RHD modeling using the RADYN code. Since the RADYN code treats scattering redistribution with a CRD approximation that has been proven to be invalid for the Mg II lines, we recalculate the Mg II line profiles from atmosphere snapshots of RADYN outputs using the RH code, which employs an angle-dependent PRD approximation \citep{Leenaarts2012}. In order to evaluate the line broadening caused by the quadratic Stark effect (i.e. pressure broadening caused by electrons and ions) more precisely, we utilize Stark full width calculation results from the STARK-B database (\url{http://stark-b.obspm.fr}) instead of the results from the current RH code.

The rest of the paper is organized as the following: Section~\ref{sec2} gives a brief introduction to the numerical codes and simulation setup. In Section~\ref{sec3}, we investigate the line broadening problem. Evolution of the atmosphere and synthetic Mg II profiles are presented in Section~\ref{sec4}. Contribution function analysis for the Mg II lines is described in Section~\ref{sec5}. We discuss the major results and their implication in Section~\ref{sec6}. Finally we present a brief summary in Section~\ref{sec7}.

\section{Method}\label{sec2}
\subsection{RHD Modeling with the RADYN Code}\label{sec2.1}
We utilize the RADYN code \citep{Carlsson1992,Carlsson1997,Allred2005,Allred2015} to perform 1D plane-parallel RHD modeling of a flare with non-thermal electron heating. A similar approach has been taken in many previous flare models (e.g., \citealt{Rubio2015,Kowalski2017a}). The code solves the time-dependent NLTE radiative transfer equation as well as hydrodynamic equations by assuming that accelerated electrons propagate downward along a 1D flux tube and deposit energy in and below the corona. Here we use a 5F11 electron beam model (maximum energy flux equals $5\times10^{11}$ erg cm$^{-2}$ s$^{-1}$) similar to the one previously adopted by \citet{Kowalski2017a}. The energy flux is similar to the energy flux inferred by \citet{Kleint2016} from the RHESSI observation of the X1.1-class flare on 2014 March 29, $3.5\times10^{11}$ erg cm$^{-2}$ s$^{-1}$. The electron beam heats the atmosphere in an extended timescale, and its energy flux ramps up and down with a FWHM duration of 20 s and a maximum flux of 5F11. This time evolution of the beam flux was employed in order to study the gradual phase of a heating pulse. Time steps in the RADYN code will be extremely small when a shock develops in the chromosphere \citep[see][]{Kowalski2015}. In order to avoid that, we modify the second derivative of the adaptive grid weights and accuracy of the minor level populations in RADYN like \citet{Kowalski2015}. Detailed treatment of radiative transfer processes makes it possible to evaluate the energy loss by radiation and produce synthetic spectra of important transitions. For a more detailed description of the setup in the RADYN code, we refer to \citet{Kowalski2017a} and \citet{Allred2015}.

\subsection{NLTE radiative transfer calculation with the RH Code}\label{sec2.2}
We use the RH code \citep{Uitenbroek2001,Pereira2015b} to recalculate 1D plane-parallel NLTE radiative transfer, with a new treatment of the Mg II quadratic Stark broadening (see Section~\ref{sec3}). The RH code takes snapshots of the RADYN output as its input. These snapshots contain height dependent information pertaining to the temperature, electron density, hydrogen population, bulk velocity and micro-turbulent velocity. The RH code recalculates the excitation and ionization populations of given atoms in statistical equilibrium, though such a condition is likely violated during the impulsive phase of flares \citep{Abbett1999,Allred2005,Rubio2017}. With certain atoms/ions treated in NLTE, the RH code can solve the radiative transfer equations for spectral lines under PRD with an angle-dependent approximation as described in \citet{Leenaarts2012}. The PRD approximation has been demonstrated to be more appropriate for the calculation of the Mg II lines that are formed by scattering.

We ran the RH code with a new evaluation of the quadratic Stark effect to solve the radiative transfer equation at a heliocentric angle of 40 degree ($\mu = 0.77$) for a direct comparison with the IRIS observations of the X1.1 flare on 2014 March 29. A six-level hydrogen atom model and a Mg II atom model with 10 energy levels are treated in NLTE in the RH code, while other atoms like Ca and He are calculated as LTE background. Scattering in the Mg II h \& k doublets and subordinate lines are evaluated with the "hybrid" angle-dependent PRD approximation \citep{Leenaarts2012}. We set the main RH convergence limit as $10^{-2}$. In addition, we cut off the corona with a temperature above 8 MK to facilitate convergence.

\section{line broadening problem}\label{sec3}
Extremely broadened profiles for chromospheric spectral lines including the above mentioned Mg II lines are generally missing in previous 1D RHD  flare simulations (e.g., \citealt{Rubio2016}). The far wings of Mg II h \& k exceeding $\pm$1.5~\mbox{\AA} ($\sim\pm$160 km s$^{-1}$) usually reveal significant enhancement in IRIS observations of X-class flares. However, the RH code often produces Mg II line profiles with little emission in the wings beyond $\pm$ 0.5 \mbox{\AA}.

\begin{figure}[ht]
    \centering
    \plotone{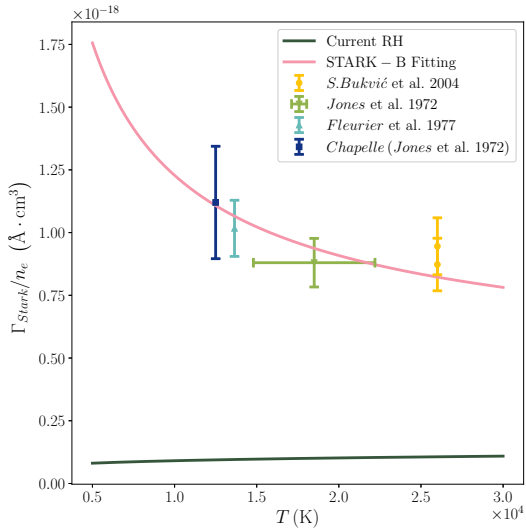}
    \caption{Comparison of Mg II h \& k full Stark width normalized to the electron density given by the STARK-B database and the current RH code. Several experimental results from \citet{Jones1972,Fleurier1977,Bukvic2004} are also included to indicate the accuracy of the  STARK-B results. The Chapelle result is mentioned in \citet{Jones1972}.\label{RH-SB}}
\end{figure}

\begin{figure*}
    \centering
    \includegraphics[width=\textwidth]{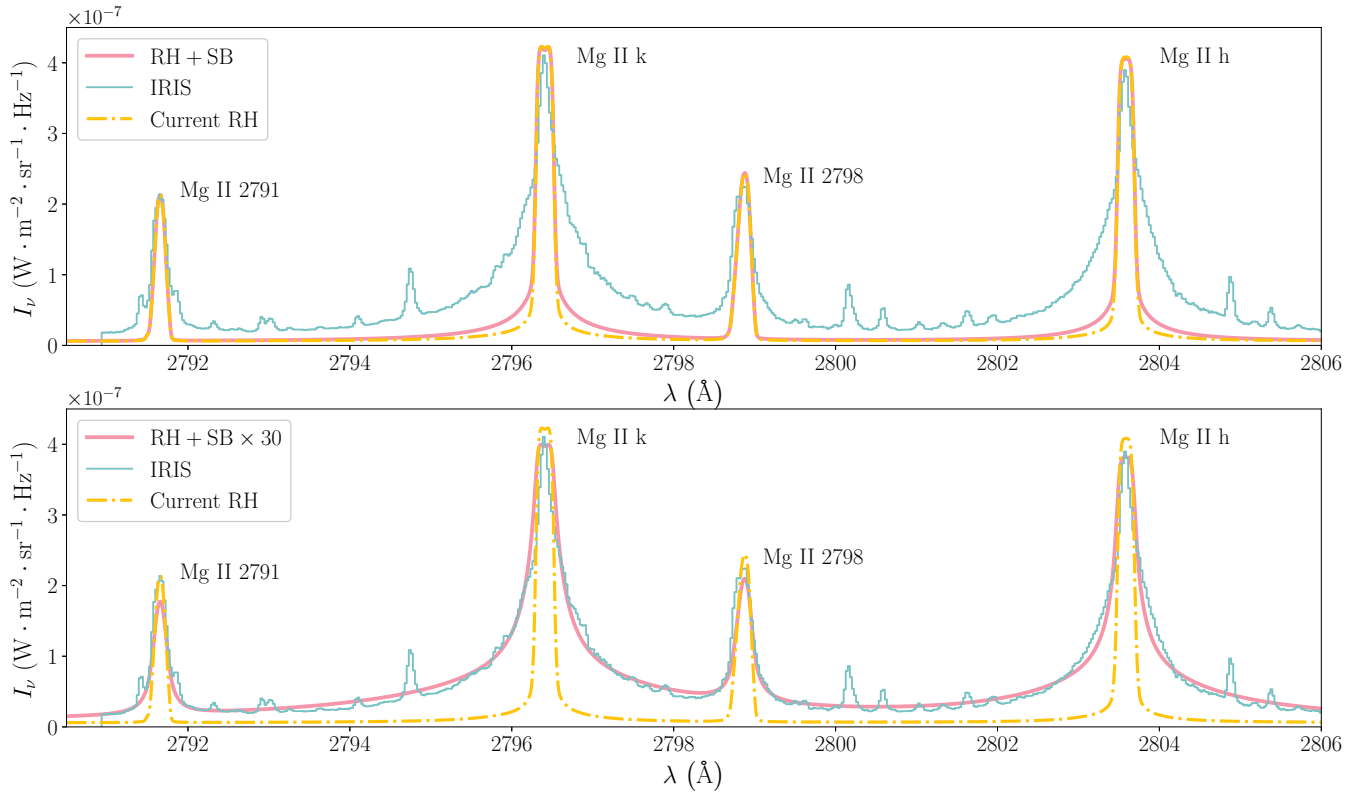}
    \caption{Comparison between the observed and synthetic Mg II profiles. In both panels the cyan and yellow curves represent the observed profiles (from ribbons of the X1.1 flare on 2014 March 29) and the synthetic profiles from the current RH code, respectively. The red curve refers to the synthetic profiles from the RH+SB calculation in the upper panel and from the RH+SB$\times$30 calculation in the lower panel.  \label{MgII-SB}}

\end{figure*}

\begin{figure*}
    \centering
    \includegraphics[width=\textwidth]{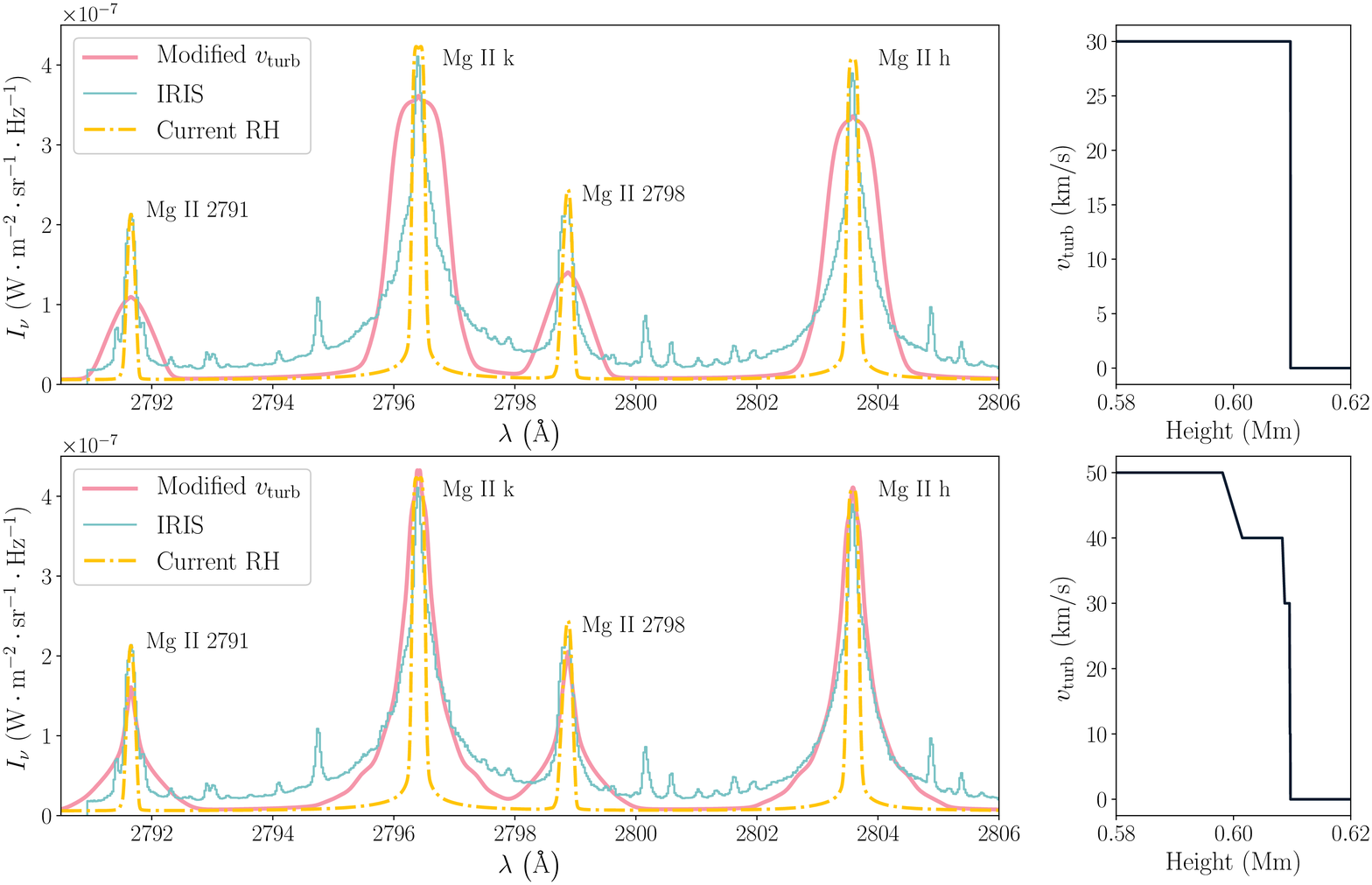}
    \caption{Synthetic Mg II profiles from the RH+SB calculation with modified micro-turbulent velocities (red curves). Upper panel: result by introducing a constant turbulent velocity of 30 km s$^{-1}$ below the h \& k line core formation height; Lower panel: result by introducing a gradually ascending turbulent velocity to 50 km s$^{-1}$ towards lower heights. In both panels the cyan and yellow curves represent the observed profiles (from ribbons of the X1.1 flare on 2014 March 29) and the synthetic profiles from the current RH code (5 km s$^{-1}$ micro-turbulent velocity at all grids), respectively. \label{MgII-vturb}}
\end{figure*}

The quadratic Stark effect is an important broadening mechanism for some chromospheric lines at flare ribbons. The heating and subsequent ionization of hydrogen increases the electron density dramatically by $\sim$1-3 orders of magnitude. A large amount of electrons and protons perturb the Mg II energy levels during scattering. As a result, more photons are scattered into the far wings. Several previous modeling efforts of the Mg II and other chromospheric lines suggested that the RH code underestimates electron pressure broadening (e.g., \citealt{Mihalas1978,Kowalski2017a,Kowalski2017b}). Two classical theories are widely used to evaluate the pressure broadening: the impact theory that results in a Lorentzian profile \citep{Weisskopf1932}, and the statistical theory (or quasi-static approximation) which leads to a Holtsmark profile \citep{Holtsmark1919}. The impact theory is more appropriate for Mg II electron pressure broadening, since half of the wavelength range ($\Delta\lambda$) corresponding to the broadened wing is much smaller than the Weisskopf radius (a $\Delta\lambda$ below which the impact theory still holds). The current RH code treats quadratic Stark broadening in classical impact theory with the Lindholm approximation \citep{Lindholm1945,Foley1946,Mihalas1978}, which gives a full Stark width of 
\begin{equation}
    \Gamma_{Stark}=11.37C_4^{2/3}v_{rel}N_e
\end{equation}
where $v_{rel}$ is the relative velocity of electron with respect to the emitter (Mg II ions), $N_e$ is local electron density and $C_4$ is the quadratic Stark effect constant. It describes the change of emitting angular frequency $\Delta \omega = C_4/r^4$ when a perturber passes at distance $r$. The value of $C_4$ is different for different transitions. In the current RH code the calculation of $C_4$ is based on the approximation of \citet{Traving1960}, which only depends on the properties of the emitting atom and the upper/lower energy levels. For the Mg II h \& k lines, the current RH code yields $C_4\approx 2.6\times10^{-16}\ \mathrm{cm^4/s}$, which is smaller than the typical range of $10^{-12}-10^{-15}\ \mathrm{cm^4/s}$ given by \cite{SobelMan1973}.

\startlongtable
\begin{deluxetable*}{cccccc}
    \tablecaption{Fitting coefficients for the full Stark width derivation}
    
    \tablenum{1}
    
    \tablehead{\colhead{Spectral Line} & \colhead{Transition} & \colhead{Wavelength} & \colhead{$a_0$} & \colhead{$a_1$} & \colhead{$a_2$} \\ 
    \colhead{} & \colhead{} & \colhead{($\mbox{{\AA}}$)} & \colhead{} & \colhead{} & \colhead{} } 
 
    \startdata
    Mg II h& $3p\ ^2\mathrm{P}_{1/2}-3s\ ^2\mathrm{S}_{1/2}$  & 2803.53 & \multirow{2}{*}{1.13807} &\multirow{2}{*}{-1.54913} & \multirow{2}{*}{0.13423}\\ 
    Mg II k& $3p\ ^2\mathrm{P}_{3/2}-3s\ ^2\mathrm{S}_{1/2}$  & 2796.35\\ 
    \hline
    Mg II 2791  & $3d\ ^2\mathrm{D}_{3/2}-3p\ ^2\mathrm{P}_{1/2}$ &2791.60& \multirow{3}{*}{1.36331}& \multirow{3}{*}{-1.62649} & \multirow{3}{*}{0.15218}\\ 
    \multirow{2}{*}{Mg II 2798}  & $3d\ ^2\mathrm{D}_{3/2}-3p\ ^2\mathrm{P}_{3/2}$ &2798.75\\
     & $3d\ ^2\mathrm{D}_{5/2}-3p\ ^2\mathrm{P}_{3/2}$ & 2798.82
    \enddata
    \label{Table1}
    \tablecomments{$\ n_e = 10^{15} \mathrm{cm^{-3}}$}
\end{deluxetable*}

There are a few preconditions for the validity of the classical impact theory, one of which is the adiabatic approximation. According to \citet{Mihalas1978}, the impact frequency $\Delta \omega_s$ should be small enough compared with any transition frequencies. Otherwise a transition occurring during the perturbation would change the energy and break the adiabatic assumption. In our case, when the perturbers are electrons, the impact angular frequency $\Delta \omega_s\sim10^{15}\ \mathrm{rad/s}$ and the Mg II h \& k transition frequency $\omega_{ij} \approx 6.7\times10^{15}\ \mathrm{rad/s}$ have the same order of magnitude. This implies a failure of the adiabatic approximation and suggests that the implementation of the impact theory for the line broadening calculation is not appropriate.  

For a better evaluation of the line broadening caused by the quadratic Stark effect, we decide not to use the current Stark width calculation in the RH code. Instead, we obtain the Mg II h, k and triplet full Stark widths from the STARK-B database (\url{http://stark-b.obspm.fr}). The widths are calculated based on an impact-semiclassical-perturbation theory \citep{Dimitrijevi1995,Dimitrijevi1998}. We use an interpolation formula given by the database (\citealt{Sahal2011}) to calculate the full Stark widths at different temperatures:
\begin{equation}
    \mathrm{Log}\left(\Gamma \left(\mbox{{\AA}}\right)\right) = a_0 + a_1\mathrm{Log}(T) + a_2\mathrm{Log}^2(T) 
\end{equation}
where $a_0, a_1$ and $a_2$ are fitting coefficients given by the STARK-B database (shown in Table~\ref{Table1}). The coefficients for the Mg II h \& k lines are the same. And the subordinate lines share the same coefficients. Note that $a_0$ depends on the electron density to ensure $\Gamma\propto N_{\mathrm{e}}$.

Figure~\ref{RH-SB} presents the temperature dependence of Mg II h \& k full Stark width normalized to electron density, as given by STARK-B and implemented in the current RH code. Obviously, STARK-B gives much larger Stark widths compared to the current RH code. The unmodified RH code gives a full Stark width $\sim$ 1 order of magnitude smaller than that from STARK-B at $T=7\times10^3-1\times10^4$ K, where the Mg II h \& k wings likely form (see Section~\ref{sec5}). In addition, as the temperature decreases, STARK-B gives an even larger Stark width as a result of increasing inelastic collisions. Figure~\ref{RH-SB} also shows several experimental results (\citealt{Jones1972}; \citealt{Fleurier1977}; \citealt{Bukvic2004}), which are consistent with the prediction from the STARK-B database. Hence, we conclude that it is more accurate to evaluate the Stark broadening of the Mg II lines by introducing results from the STARK-B database. We thus implement the STARK-B result in the RH code and perform the radiative transfer calculation (referred as the RH+SB calculation).

We perform both the RH+SB calculation and the current RH calculation using the same RADYN 5F11 snapshot as input. Results are shown in the upper panel of Figure~\ref{MgII-SB}. An observed IRIS spectrum at ribbons of the X1.1 flare on 2014 March 29 is also presented for comparison. After radiometric calibration, the spectra have been multiplied by a factor of 36 for the purpose of comparison with the synthetic spectra. This factor is likely related to the spatial resolution or filling factor. We found that a smaller factor of 10 was used by \citet{Rubio2017}, which is related to the fact that they took the observed spectra from some brighter pixels for comparison. Compared to the Mg II line profiles calculated from the current RH code, the profiles from the RH+SB calculation are more significantly enhanced at the far wings. However, the broadening at the far wings is still much smaller than the observation. Note that the entire line profile is a convolution of a Gaussian profile (due to thermal motion and micro-turbulence) and a Lorentzian profile (due to natural and pressure broadening). Though the full Stark width increases by about 1 order of magnitude, the line cores and near wings are still dominated by the Doppler broadening caused by thermal motion and micro-turbulence. We also notice that the triplets from the RH+SB calculation show no obvious difference compared to those from the current RH calculation.

To test how much Stark full width (i.e. Lorentzian width) is required to reproduce the significantly enhanced line wings, we have made several attempts by simply multiplying the STARK-B Stark widths by different factors and recalculating the Mg II line profiles. We find that the Stark widths from the STARK-B database have to be enlarged by 30 times (referred as RH+SB$\times$30) in order to reproduce an extremely broadened Viogt profile that fits the IRIS observation. As we can see from the lower panel of Figure~\ref{MgII-SB}, the spectrum from the RH+SB$\times$30 calculation matches the observed one surprisingly well in not only the line cores and near wings, but also the far wings and continuum.

One physical process that might be responsible for the exceeding broadening is micro-turbulence, which could increase the Gaussian component in a Voigt profile. To examine this, we insert turbulent velocities at different heights manually and calculate the synthetic spectra using the RH+SB code. We find that the broadenings of the synthetic line profiles are highly dependent on manually introduced micro-turbulent velocities at a few grids. We show the synthetic Mg II spectra in an atmosphere with a sudden increase of the turbulent velocity to $\sim$30 km s$^{-1}$ below the line core formation height (inferred from the contribution function analysis in Section~\ref{sec5}) in the upper panel of Figure~\ref{MgII-vturb}. All the Mg II lines become much more enhanced around the line cores, which are unrealistic. When we introduce  gradually changed turbulent velocities within a thin layer, sufficient broadening of the Mg II h \& k lines appears to be reproduced (shown in the lower panel of Figure~\ref{MgII-vturb}). However, this also produces much more broadened triplets compared to the observation. Also, the existence of micro-turbulence with an amplitude more than 10 km s$^{-1}$ in the chromosphere appears to be unlikely. For instance, \citet{Rodrguez2016} found a value of 8 km s$^{-1}$ in their NLTE inversion of Mg II lines. \citet{Carlsson2015} found that the Mg II k width is sensitive to the chromospheric temperature and turbulent velocity. They found a best fit value of $\sim$ 7 km s$^{-1}$ for the turbulent velocity in plage. Based on the above consideration, we can rule out micro-turbulent velocity as the cause for the factor of 30. \citet{Rubio2017} performed a similar analysis with the current RH code and reached the same conclusion.

The fact that a good match is achieved by simply multiplying the STARK-B Stark widths by 30 suggests that the significantly enhanced far wings of Mg II are likely caused by a certain physical mechanism that produces a Lorentzian profile (e.g,. pressure broadening caused by electrons or Van der Waals interaction). We rule out Van der Waals broadening because it should not differ too much at flare ribbons and in the quiet Sun. Also simulations of the Mg II lines in the quiet sun (e.g., \citealt{Uitenbroek1997,Leenaarts2013a,Rodrguez2016}) show no significant underestimation in Van der Waals broadening.

\section{Atmospheric response and evolution of synthetic line profiles}\label{sec4}

\subsection{Atmosphere Evolution}\label{sec4.1}

\begin{figure*}[ht]
    \centering
    \includegraphics[width=0.6\textwidth]{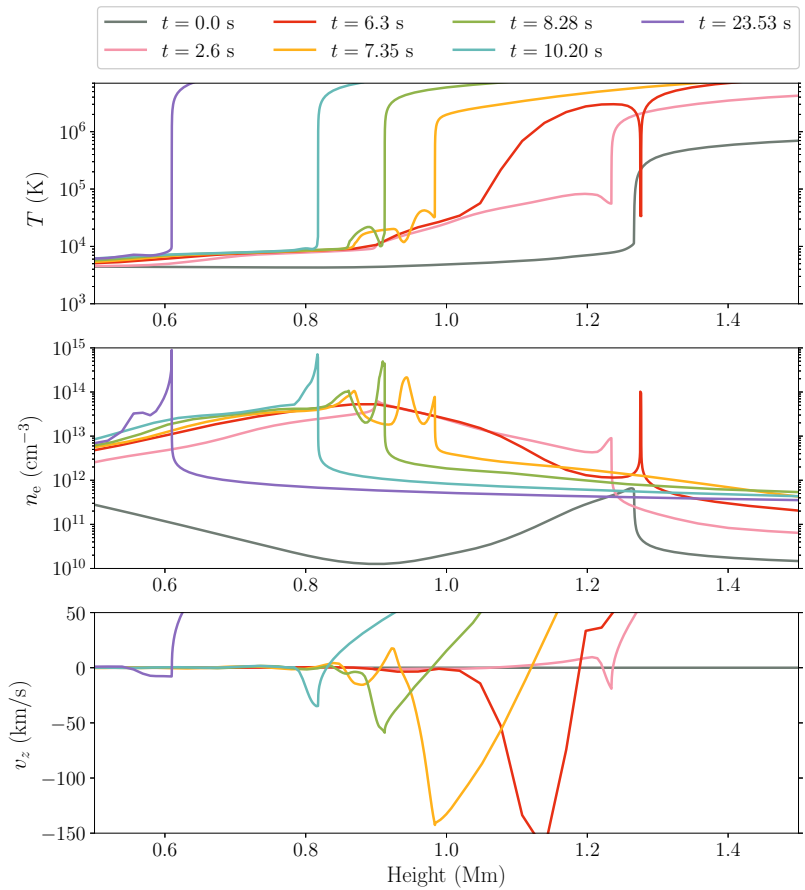}
    \caption{Atmosphere evolution in the 5F11 model. Curves in different colors represent different times ($t$) in the simulation. Panels from top to bottom: temperature, electron density and 1D velocity as a function of height. Positive and negative velocities represent upflows and downflows, respectively. Only the upper chromosphere, TR and lower corona are shown.\label{5F11_atmos}}
\end{figure*}

\begin{figure*}
\gridline{\fig{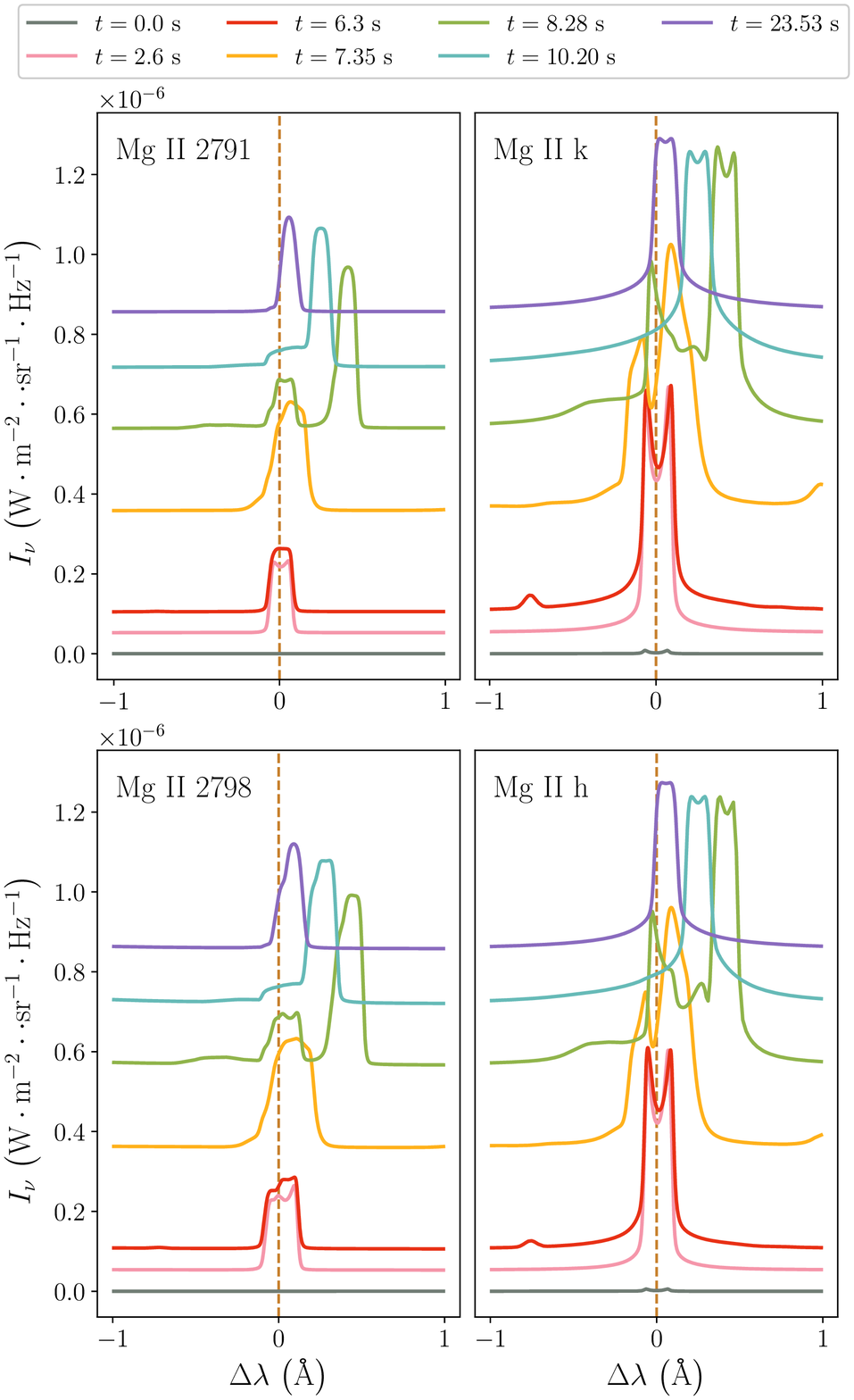}{0.5\textwidth}{(a)}
          \fig{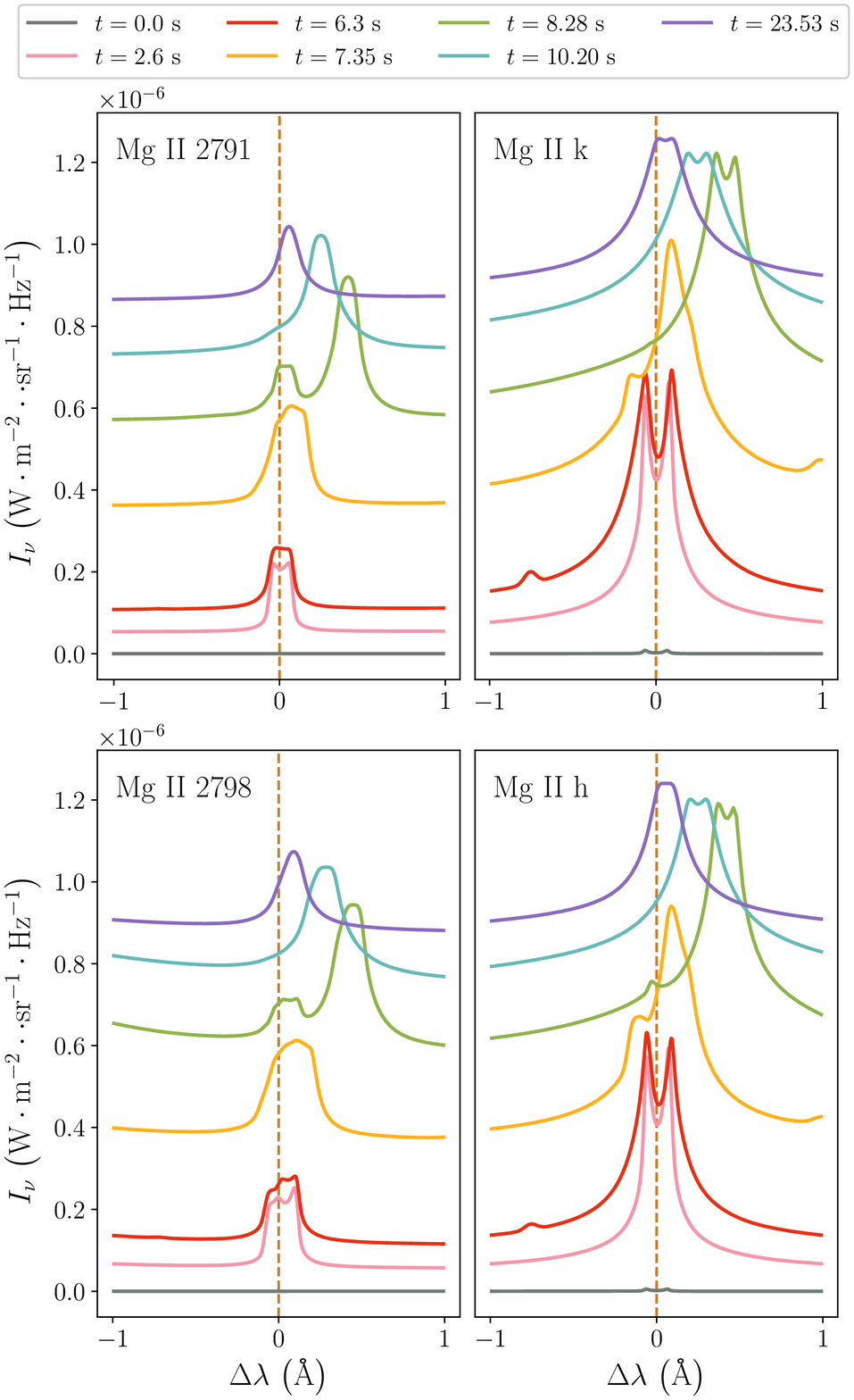}{0.5\textwidth}{(b)}}
\caption{Temporal evolution of the synthetic Mg II spectra. Different colors represent different times ($t$) in the simulation. Left: result from the RH+SB calculation. Right: result from the RH+SB$\times$30 calculation. The spectral profiles at different times are shifted in the y axis just for the purpose of illustration. \label{5F11_spectra}}
\end{figure*}

Figure~\ref{5F11_atmos} demonstrates how the atmosphere evolves in the 5F11 model. The highly dynamic atmosphere in the simulation is a direct result of extended heating by electron beams. Here we mainly focus on the dramatic changes of physical quantities such as the temperature and electron density in the upper chromosphere, where the Mg II lines form. 

At $t=2.6$ s, the upper chromosphere is significantly heated to more than $3\times10^4$ K. The ionization of neutral hydrogen and helium greatly increases the local electron density by 1--3 orders of magnitude to $10^{12}-10^{13}\ \mathrm{cm^{-3}}$. An upflow component with a speed up to $\sim$200 km s$^{-1}$, which clearly results from chromospheric evaporation, occurs above $\sim1.25$ Mm. Meanwhile, a small downflow component with a $\sim$20 km s$^{-1}$ speed forms below the bottom of the TR. The TR also shifts to a lower height by $\sim$ 25 km compared to its initial location. 

After another 3.7 s heating, the tiny downflow grows dramatically and reaches a maximum velocity of more than 150 km s$^{-1}$ in the TR. This downflow spans a range of $\sim$100 km in height. It smooths the steep temperature increase as well as electron density decrease in the TR, resulting in an extended TR spanning from $\sim$1.0 Mm to $\sim$1.2 Mm. A cold and dense shock propagates in the lower corona at $t=6.3$ s, and we smooth it later to enable reasonable time steps in RADYN code. 

At $t=7.35$ s, the $\sim$150 km s$^{-1}$ downflow reaches the boundary between the lower TR and upper chromosphere. The interaction between such a strong downflow and the dense chromosphere changes the physical conditions in the upper chromosphere dramatically. Tiny bulges in temperature ($\sim1-2\times 10^4$ K) and electron density ($\sim10^{14}\ \mathrm{cm^{-3}}$) appear in the upper chromosphere. An upflow with the maximum velocity of 50 km s$^{-1}$ (at $t=7.35$ s it is only $\sim20$ km s$^{-1}$) forms just below the strong downflow. The previously stretched TR becomes compressed again, and keeps moving to lower heights. At $t=8.28$ s, some bulges of electron density merge due to downward motion of the TR, leaving a density peak of $\sim 5\times10^{14}\ \mathrm{cm^{-3}}$ just below the TR. Meanwhile, the downflow velocity is reduced to $\sim$50 km s$^{-1}$ as it interacts with the dense chromosphere. 

At $t=10.20$ s and $t=23.53$ s, the TR still moves downward and compresses the chromosphere, though the downflow velocity is only $5-40$ km s$^{-1}$. Such a strong compression smooths out all the previously existing tiny substructures like the bulges in the temperature and electron density curves. The atmosphere becomes less dynamic. Compared to the initial atmosphere, the chromosphere is greatly compressed to lower heights and heated by a few thousand Kelvin. The electron density increases by 2--3 orders of magnitude to $5\times10^{14}-10^{15}\ \mathrm{cm^{-3}}$ just below the TR, which is close to the electron density required to reproduce non-reversed Mg II h \& k profiles in a previous parameter study \citep{Rubio2017}.  

\subsection{Temporal Evolution of Mg II Line Profiles}\label{sec4.2}

Figure~\ref{5F11_spectra} shows the temporal evolution of the Mg II spectra from the RH+SB (a) and RH+SB$\times$30 (b) calculations. The Mg II h, k, 2791 \mbox{\AA} and 2798 \mbox{\AA} profiles at different time steps are shown in different colors. In the following we examine the temporal changes of the Doppler shift and line wing emission of all these Mg II lines, as well as the central reversal of the h and k lines. Though the radiative transfer equation is solved in the same atmosphere, far wing emissions of all the Mg II lines have been greatly enhanced in the RH+SB$\times$30 calculation. At some time steps like $t=7.35$ s and $t=8.28$ s, significantly enhanced line wings even smooth out some multi-peak features.

At $t=2.6$ s and $t=6.3$ s, the Mg II h \& k lines are greatly enhanced as a result of the chromospheric heating by non-thermal electrons. The reversed line core is caused by the decoupling of the source function and Planck function at the line core formation height, where the electron density is not large enough to maintain LTE \citep{Rubio2016}. The source function decreases with height across the line core formation layer. Since the Stark broadening usually influences the scattering in far wings, line profiles from the RH+SB$\times$30 calculation also show central-reversed line cores. A tiny emission enhancement is located at the far blue wing of $\sim0.9$ \mbox{\AA} at $t=6.3$ s. As a result of chromospheric heating, the triplet lines turn from absorption to emission at these occasions. 

At $t=7.35$ s, both the h \& k lines and triplets become much more broadened compared with those at previous times. The enhanced broadening is caused by the superposition of large downflows and upflows within one model loop and the increased electron density during condensation. In the RH+SB calculation, the triplets are redshifted by $\sim 10$ km s$^{-1}$ and the Mg II h \& k lines have asymmetric spectral profiles. The red wing peak ($\mathrm{k_{2R}}$ or $\mathrm{h_{2R}}$) is much stronger than the blue wing peak ($\mathrm{k_{2V}}$ or $\mathrm{h_{2V}}$). However, in the RH+SB$\times$30 calculation, the enlarged Stark broadening scatters more photons at $\mathrm{k_{2R}}$ and $\mathrm{h_{2R}}$ into the central-reversed line cores. Consequently, the central-reversed features at the Mg II h \& k line cores are largely smeared out. 

At $t=8.28$ s, all synthetic Mg II line profiles from the RH+SB calculation consist of a stationary component and a significantly redshifted component. The multiple peaks obviously result from the complex atmospheric structures at this time. Similarly, in the RH+SB$\times$30 calculation, strong scattering at the far wings smooths these features in the h \& k lines, resulting in a single component with a red shift of $\sim$50 km s$^{-1}$. However, the triplets still reveal a clearly-separated two-component profile, which is rarely seen in observations. The greatly redshifted component may contribute to the formation of asymmetric Mg II line profiles at the moving front of flare ribbons \citep[See][]{Panos2018}.

At $t=10.20$ s and $t=23.53$ s, the compression of chromosphere and simple temperature structure below the TR lead to single-component profiles for the h \& k lines and triplets. The red shifts of these lines decrease with time as a consequence of decreasing downflow velocity from $\sim$50 km s$^{-1}$ to $\sim$5 km s$^{-1}$. With an increase of electron density below the TR, the central reversals of Mg II h \& k become less and less prominent. Eventually they almost disappear. The RH+SB$\times$30 calculation yields much more broadened line profiles and reveals no other significant difference.

\section{contribution function analysis}\label{sec5}

\begin{figure*}
    \gridline{\fig{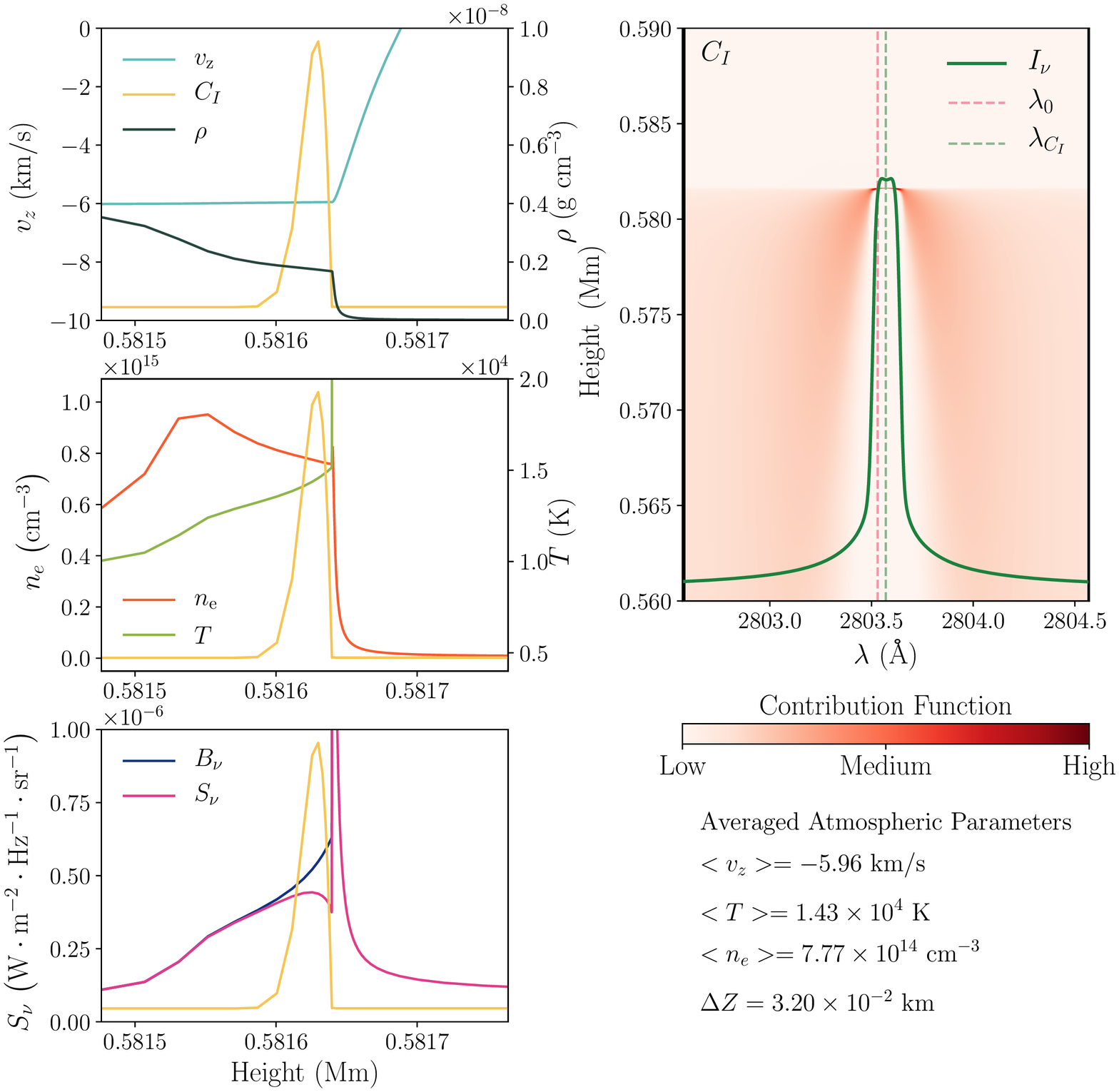}{0.5\textwidth}{(a)}
          \fig{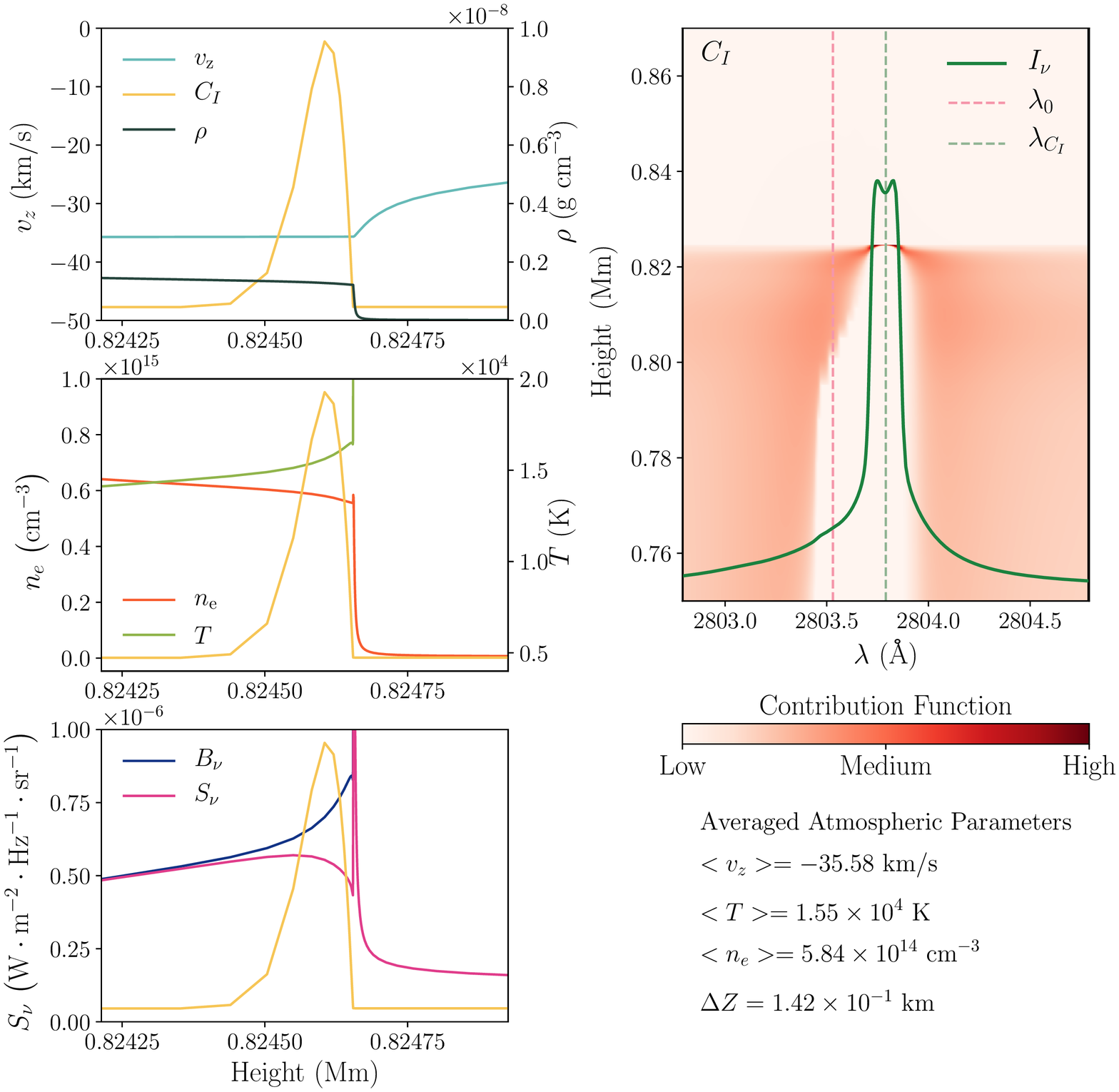}{0.5\textwidth}{(b)}}
    \gridline{\fig{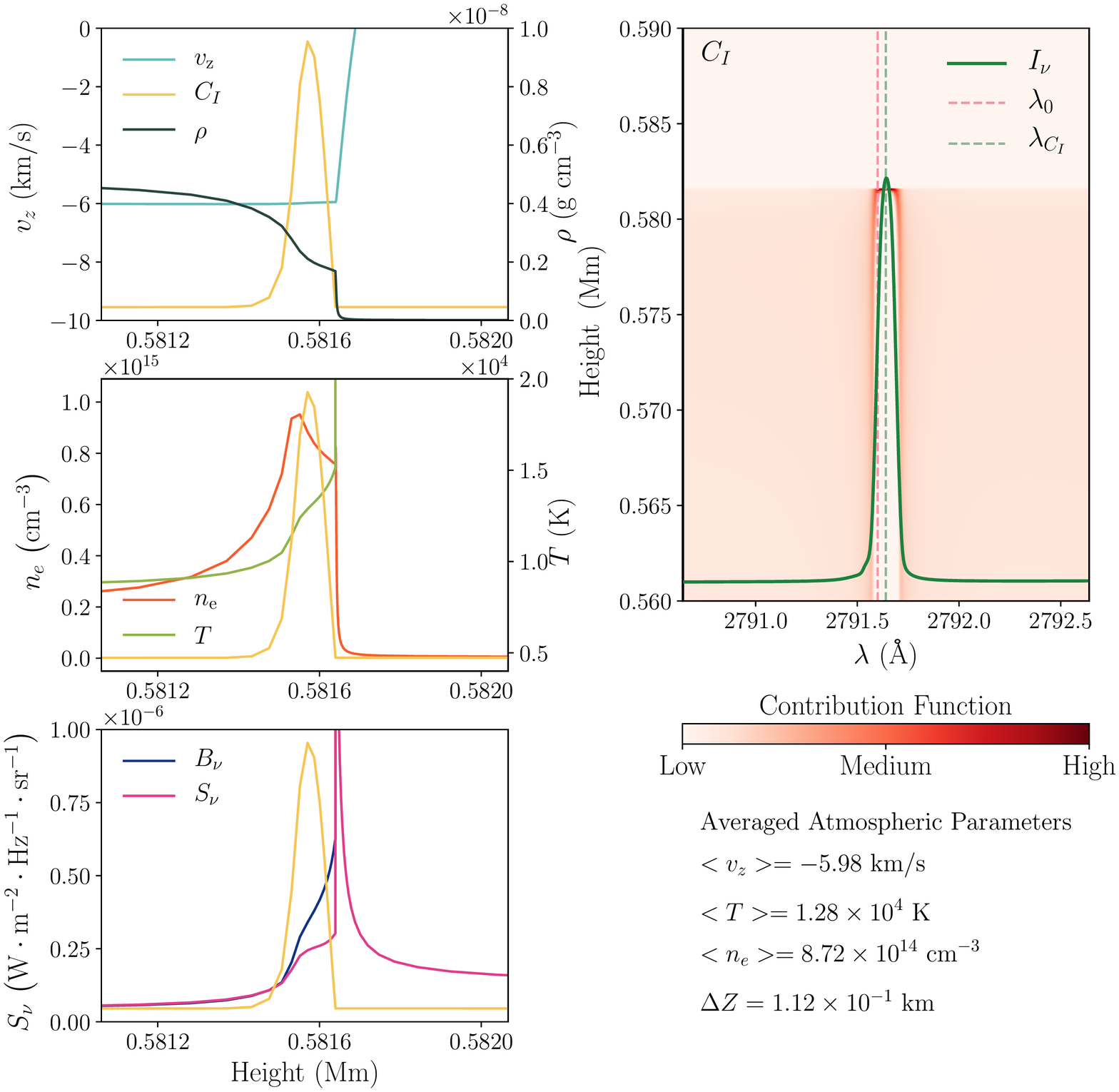}{0.5\textwidth}{(c)}
          \fig{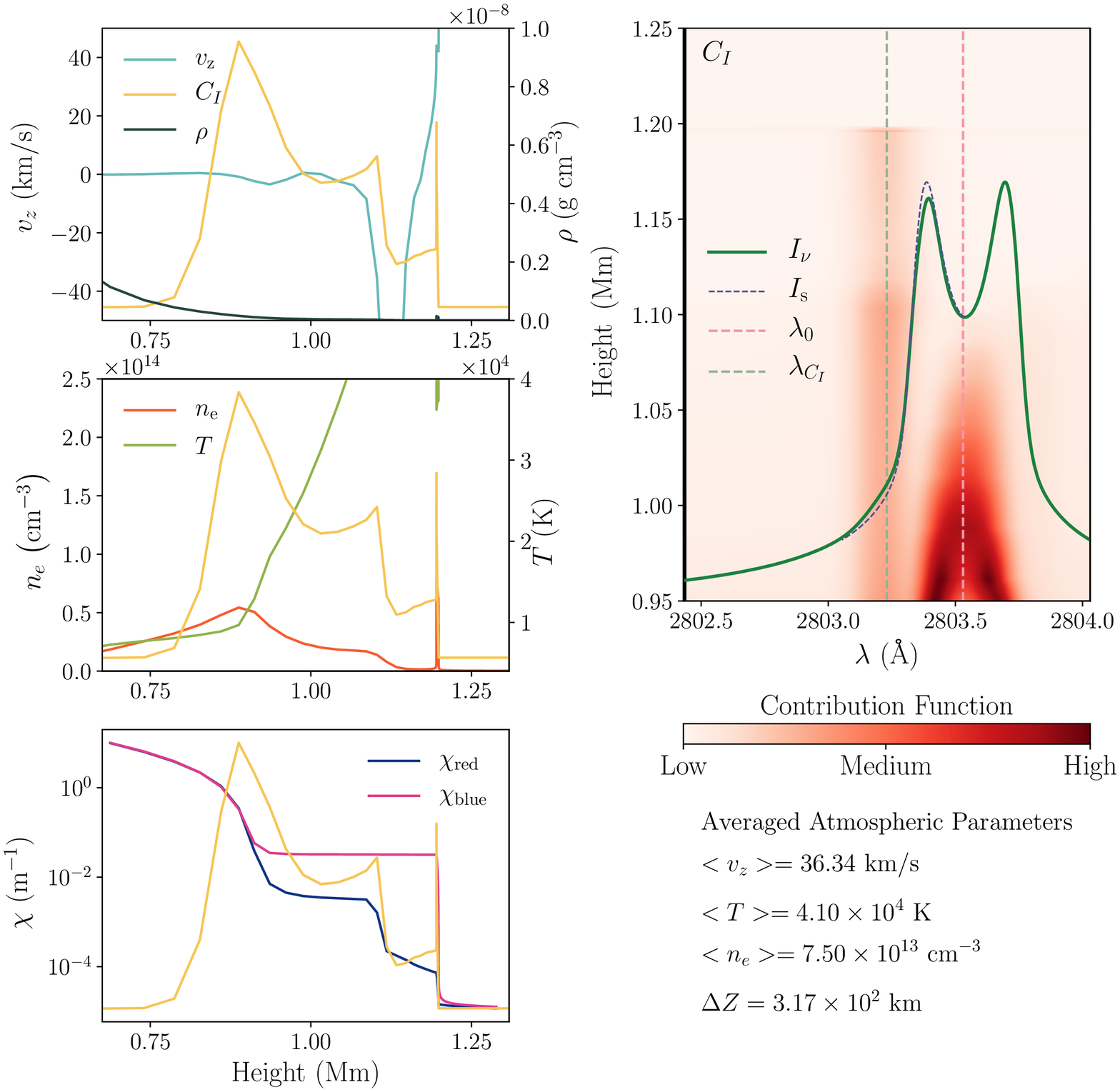}{0.5\textwidth}{(d)}}
    \caption{RADYN atmospheric snapshots and distribution of the contribution functions of the Mg II lines. Panel (a), (b) and (c) are calculated in RH+SB while (d) is calculated in RH+SB$\times$30. (a) Mg II h line profile at $t=27.56$ s. Left panels: 1D velocity, mass density, electron number density, temperature, source function and Planck function. Right panel: distribution of the contribution function at different wavelengths and heights. The weighted average of velocity, electron density and temperature, as well as the thickness of line core formation region are printed in the lower right corner. The green solid line represents the spectral line profile $I_{\nu}$. The red and green dotted lines mark the rest wavelength $\lambda_0$ and the wavelength ($\lambda_{C_I}$) at which the contribution function curve ($C_I$) is plotted in the left panels, respectively. (b) Same but for the Mg II h line at $t=10.0$ s. (c) Same but for the Mg II 2791 \mbox{\AA} line at $t=27.56$ s. (d) Similar but for the Mg II h line at $t=5.30$ s from the RH+SB$\times$30 calculation. The lower left panel shows the height variations of opacity at the blue wing ($-0.3~$\mbox{\AA}) and red wing ($+0.3~$\mbox{\AA}). Note that in (d) the atmospheric parameters are averaged around the upflow rather than over the entire range of formation height. A symmetric line profile $I_{s}$, which is extrapolated by using the red half of $I_{\nu}$, is shown as the blue dashed curve. \label{5F11_ctbfun1}}
\end{figure*}

\begin{figure}
    \plotone{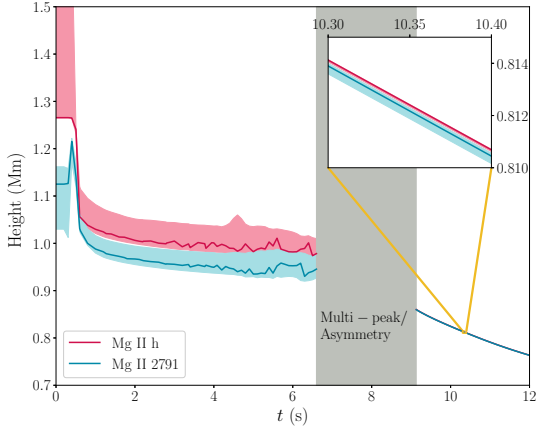}
    \caption{Temporal evolution of the Mg II h and 2791 \mbox{\AA} line core formation heights. The red and cyan colors represent results of Mg II h and the 2791 \mbox{\AA} triplet, respectively. For each line the shaded region stands for the line core formation layer (only shows the height range where $C_I'$ is between 0.16 and 0.84). The data between $t=6.80$ s and $t=9.13$ s are not displayed, since the highly asymmetric or multi-peak line profiles make it difficult to determine the line core wavelength and formation height during this period. The solid line indicates the height where the contribution function peaks. An inset showing how close the Mg II h and 2791 \mbox{\AA} cores form is also presented in the upper right corner. \label{formation_height}}
\end{figure}

In order to have a better understanding of the formation of the Mg II lines, we also perform an analysis of the contribution function for each line. The contribution function is defined as the following \citep{Magain1986,Carlsson1997,Kowalski2015}:
\begin{equation}
    C_I = \frac{\mathrm{d}I_{\nu}}{\mathrm{d}z}=\frac{1}{\mu}\chi_{\nu}S_{\nu}e^{-\tau_{\nu}/\mu}
\end{equation}
where $\chi_{\nu}$ is the opacity, $\tau_{\nu}$ is the optical depth integrated from $\chi_{\nu}$ along height. $S_{\nu}$ is the source function defined as the ratio between emissivity $j_{\nu}$ and opacity $\chi_{\nu}$. These quantities are all functions of frequency $\nu$. To determine the height where most emission comes from, we also calculate the cumulative contribution function defined in \cite{Kowalski2016}:
\begin{equation}
    C'_I(z,\mu) = 1- \frac{\int^{z=10\mathrm{Mm}}_{z\ge z_{\mathrm{lim}}} C_I(z,\mu)\ \mathrm{d}z}{\int^{z=10\mathrm{Mm}}_{z = z_{\mathrm{lim}}} C_I(z,\nu)\ \mathrm{d}z}
\end{equation}
In addition, we define the formation layer of the emission at frequency $\nu$ between $z\left(C'_I=0.95\right)$ and $z\left(C'_I=0.05\right)$. The thickness of the formation layer $\Delta Z$ can then be calculated. We can also calculate the weighted average of a physical quantity as a probe of the local atmospheric conditions within the formation layer:
\begin{equation}
    <X> = \frac{\int^{z\left(C'_I=0.95\right)}_{z\left(C'_I=0.05\right)}C_I(z,\mu)X\ \mathrm{d}z}{\int^{z\left(C'_I=0.95\right)}_{z\left(C'_I=0.05\right)}C_I(z,\mu)\ \mathrm{d}z}
\end{equation}
where X could be temperature $T(z)$, electron density $n_e(z)$ or velocity $v_z(z)$.

Figure~\ref{5F11_ctbfun1}a demonstrates how a typical non-reversed Mg II h line forms at $t=27.56$ s in the 5F11 model. The contribution function indicates that the virtually non-reversed line core is formed in an extremely thin layer ($\Delta Z\approx$ 32 m), which is well resolved by the adaptive grid. The source function starts to deviate from the Planck function at the line core formation height. An average electron density of $7.77\times10^{14}\ \mathrm{cm^{-3}}$ and a continuous increase of temperature across the line core formation region result in a continuous increase followed by a slight decrease in the source function. The line profile is also slightly redshifted, corresponding to an average downward velocity of $5.96$ km s$^{-1}$. This red shift obviously results from the compressed chromospheric plasma, which has cooled over time and decreased in speed \citep[e.g.,][]{Graham2015,Tian2018}. Compared to the line core, the far wings are formed in a more extended height range from $\sim$0.565 Mm to $\sim$0.582 Mm. 

A slightly reversed Mg II h line profile is found at $t = 10.0$ s (Figure~\ref{5F11_ctbfun1}b), when the chromosphere is not so much compressed compared to $t=27.56$ s. The line core forms in a narrow region of $\approx142$ m centered at the height of $\sim0.825$ Mm. The average downflow velocity in the line core formation layer is 35.58 km s$^{-1}$, which shifts the line by $\sim0.25$ \mbox{\AA}. The average electron density is $5.84\times 10^{14}\ \mathrm{cm^{-3}}$. Compared to $t=27.56$ s, the source function drops much faster with increasing height across the line core formation region. Hence, there are not enough photons scattered at the line core and a central reversal forms.

Figure~\ref{5F11_ctbfun1}c shows the Mg II 2791 \mbox{\AA} line contribution function at $t=27.56$ s. The Mg II 2791 \mbox{\AA} line core forms only hundreds of meters below the Mg II h core formation height. Compared to the h line core, the 2791 \mbox{\AA} line core is formed in a larger range of height ($\approx112$ m), extending downward to the cooler region with higher electron densities. Though the source function has already started to decouple from the Planck function, it still keeps increasing with height and leads to a line core in emission.

Figure~\ref{5F11_ctbfun1}d displays a Mg II h line profile with slight blue wing enhancement at $\sim-0.3\ \mathrm{\AA}$, which is found at $t=5.30$ s. Here we analyze the results from the RH+SB$\times$30 calculation with a micro-turbulent velocity of 7 km s$^{-1}$ as the profiles resulted from this calculation show more emission in the far wings, similar to IRIS observations. The intensity of $\mathrm{k_{2V}}$ is slightly smaller than that of $\mathrm{k_{2R}}$ at this time. The contribution function at $\sim-0.3\ \mathrm{\AA}$ is distributed mainly from $\sim0.75$ Mm to $\sim1.20$ Mm, indicating an extended line wing formation region. The enhancement in the blue wing is related to a cool upflow with a velocity of $\approx36$ km s$^{-1}$ in the TR. These $\sim4\times10^4$ K cool materials increase the opacity at the blue wing by $\sim$1--2 orders of magnitude, leading to an elevated contribution function between 0.95 Mm and 1.2 Mm. This blue wing enhancement continues to move further to the blue in our simulation when the upflow propagates upward, since the upflow velocity is larger at higher layers.  

\section{discussion}\label{sec6}
In Section~\ref{sec3} we have successfully reproduced the extremely broadened Mg II line profiles with Lorentzian wings by multiplying the Stark widths extracted from the STARK-B database by a factor of 30. After ruling out the possibility of micro-turbulent broadening and Van der Waals broadening, we discuss several other effects which may contribute to this factor:
\begin{enumerate}
    
    \item Non-thermal electrons:  Non-thermal electrons will also interact with the Mg II ions and change the redistribution of photons in scattering like thermal electrons, resulting in additional quadratic Stark broadening \citep{Hawley2007}. Non-thermal electrons may also interact with ambient electrons first, and these energized ambient electrons will perturb Mg II ions again on their way to thermalization. In addition, the non-thermal excitation and ionization will change the population of different energy levels in the Mg II ion, which may also enhance the line broadening.  
    
    \item Three-dimensional effects: The 3D radiative transfer process has been suggested to affect the Mg II line core formation in the quiet Sun \citep{Leenaarts2013a}.  The 3D effect has not been investigated in flares, and it may contribute to the broadening of the Mg II lines.  
    
    \item Magnetic fields: The magnetic field can affect the Mg II energy levels through Zeeman effect. Previous studies have shown that the Mg II h\&k lines are sensitive to Zeeman effect even for weak fields \citep[e.g.,][]{Alsina2016}. By manually introducing a magnetic field with a strength up to 1000 Gauss in the RH code, we find that the Mg II Zeeman broadening in the Stokes \textit{I} profiles is negligible. However, it is unclear whether effects of the magnetic field will cause substantial broadening in the realistic 3D case.
    
    \item Electron density effect: \citet{Rubio2017} found that non-reversed h \& k line profiles can be reproduced by manually increasing the electron density by a factor of more than 10 to $\sim10^{15}\ \mathrm{cm^{-3}}$. In the later phase of our simulation the Mg II line cores form in a layer of similar electron density. However, the far wings form in an environment where the electron density is one order of magnitude lower. A significant increase of electron density in the lower atmosphere would greatly enhance the quadratic Stark broadening at the far wings universally in both the h \& k and triplet lines. 
      
\end{enumerate}

\citet{Rubio2017} suggested that the superposition of unresolved upflows and downflows with speeds up to 200 km s$^{-1}$ in the chromosphere can give rise to quasi-symmetric, non-reversed and extremely broadened Mg II line profiles. Also, the superpositon of two large downflows can lead to highly asymmetric Mg II line profiles with red wing enhancement extending up to 1 \mbox{\AA}. In our simulation, we also find strong downflows with speeds up to 150 km s$^{-1}$ and a $\sim$50 km s$^{-1}$ upflow when the downflows hit the upper chromosphere. The co-existence of bidirectional flows does result in several broadened profiles, i.e., at $t=7.35$ s in Figure~\ref{5F11_spectra}. However, due to the deceleration by the denser lower atmosphere, the downflow velocity in the chromosphere rarely exceeds 100 km s$^{-1}$. As a consequence, though we find a few Mg II profiles with a component that is redshifted by $0.6-0.7$ \mbox{\AA} between $t=7$ s and $t=8.5$ s, the redshifted component in the 5F11 model is incapable of producing enough far wing emission up to $+1.0$ \mbox{\AA}. Such high-speed downflows were found by \citet{Rubio2017}, and they can be triggered by electron beams with a higher energy flux like in the F13 models (e.g., \citealt{Kowalski2015,Kowalski2018}).

\citet{Panos2018} used machine learning to characterize different types of Mg II line profiles observed at flare ribbons. They found Mg II h \& k profiles with red wing enhancement at the ribbon front. After the chromosphere is swept by the ribbon front, single-peaked profiles without central reversal can be identified. The entire evolution of Mg II line profiles is similar to that in our 5F11 simulation. In our simulation, significantly redshifted profiles are seen between $t\sim7$ s and $t \sim 10$ s as a result of the large downflow. And single-peaked profiles appear after $t\sim20$ s, when the chromosphere is compressed and the electron density at the line core formation height reaches $\sim8\times10^{14}\ \mathrm{cm^{-3}}$.

In a flare observation, \citet{Tei2018} found Mg II h profiles with blue wing enhancement lasting for $\sim40$ s. The profiles then changed into those with significant red wing enhancement. They argued that the blue wing enhancement is caused by upward propagating cool materials. They performed a cloud modeling to fit the parameters of this cool cloud, and found a best-fit velocity of 40 km s$^{-1}$ after assuming a temperature of $10^4$ K. We also found a tiny blue wing enhancement in our simulation from $t\sim5.3$ s to $t\sim6.4$ s, as shown in Figure~\ref{5F11_spectra} and Figure~\ref{5F11_ctbfun1}d. This blue wing enhancement is contributed by additional emission roughly from 0.95 Mm to 1.2 Mm due to the opacity increase related to an upward propagating cool upflow ($\sim4\times10^4$ K) with a speed of $\sim$36 km s$^{-1}$. Similarly, in our simulation the blue wing enhancement occurs before the appearance of a strong redshifted component, which results from the arrival of the strong downflow at the chromosphere. Though the temperature is much higher and the lifetime is much shorter in our simulation, our result strongly supports their interpretation of the blue wing enhancement being caused by a cool upflow.

Previous 3D RMHD simulations of the quiet sun found that the central-reversed core $\mathrm{k_3}$ or $\mathrm{h_3}$ is formed only $\sim$200 km below the TR \citep{Leenaarts2013b}. They also found that the emission peaks $\mathrm{k_2}$ or $\mathrm{h_2}$ form at a height of $\sim1.4$ Mm. The Mg II triplets are believed to form in the lower chromosphere in the quiet sun \citep{Pereira2015a}. However, \citet{Rubio2017} found that triplets can form between 0.83 Mm and 1.14 Mm in the upper chromosphere during flares. In Section~\ref{sec5} we have argued that the formation heights of triplets are very close to those of the h \& k lines when the chromosphere is significantly compressed. We show the temporal evolution of the line core formation heights of Mg II h and 2791 \mbox{\AA} from the RH+SB calculation in Figure~\ref{formation_height}. The formation heights of both line cores become very close to each other after 1 s, when the electron beam heating starts. The formation regions of both line cores then move to lower heights. At $t\sim4$ s, the two formation regions overlap and become closer and closer. As the downflow velocity decreases to less than 50 km s$^{-1}$, both line cores form in extremely narrow regions separated by only a few km. The formation of triplets at the upper chromosphere suggests that properties of the triplets are determined by the upper chromospheric condition rather than the lower \citep[also see Figure 11,][]{Kerr2019}. The triplets in emission are still a signal of the lower chromosphere being heated. However, one should be cautious to use these triplets as quantitative diagnostics of heating in the lower chromosphere, because their dominant emission comes from the upper chromosphere during flares.

\section{conclusion}\label{sec7}
We have performed 1D RHD modeling of the Mg II h \& k and triplet lines at flare ribbons with the RADYN and RH codes. We find that the current RH code significantly underestimates the Mg II Stark broadening. To evaluate the Stark broadening more accurately, we have modified the RH code by implementing semi-classical perturbation approximation results from the STARK-B database (referred to as RH+SB). Compared with the results from the current RH code, the STARK-B database gives a full Stark width that is $\sim1$ order of magnitude higher and shows good consistency with lab experiments. However, the improved Stark widths have to be multiplied by a factor of 30 to reproduce the significantly broadened Mg II line profiles observed at flare ribbons. We argue that non-thermal electrons, magnetic fields as well as 3D and electron density effects may contribute to this factor.

The 5F11 model and the RH+SB calculation have reproduced a wide range of Mg II behavior seen in IRIS observations. For instance, from $t\approx5.3$ s to $t\approx6.4$ s, a weak blue wing enhancement is observed in the Mg II h \& k profiles. This blueward asymmetry is caused by a cool ($\sim4\times10^4$ K) upflow, confirming the suggestion made by \citet{Tei2018}. A strong downflow with speed of $\sim150$ km s$^{-1}$ forms at this time in the TR and propagates to the lower atmosphere. The interaction between this downflow and the dense chromosphere, as well as continuous electron beam heating, gives rise to a $\sim$50 km s$^{-1}$ upflow. The superposition of the upflow and downflow produces greatly broadened profiles at $t\approx7.35$ s. As the large downflow propagates to the lower atmosphere, the Mg II profiles show a component that is significantly redshifted by up to $\sim0.6-0.7\ \mathrm{\AA}$, which may contribute to the observed red wing enhancement of the Mg II h \& k lines.

The single-peaked Mg II h \& k profiles are found after $t\approx20$ s, when the electron beam flux and downflow velocity are both decreasing. Compared to the parameter study by \citet{Rubio2017}, we do not manually change the atmospheric parameters. So this is the first time that non-reversed Mg II h \& k profiles are reproduced in a self-consistent atmosphere. The compressed chromosphere and TR lead to an enhanced electron density of $\sim$$10^{15}\ \mathrm{cm^{-3}}$ in the upper chromosphere, maintaining an increasing source function with height in the line core formation region. 

We have also found that the Mg II h \& k lines and triplet lines are both formed in the upper chromosphere at flare ribbons. Thus the triplets are not a good quantitative diagnostic of heating in the lower chromosphere. The positive correlation between the core-to-wing ratio and temperature increase in the lower chromosphere, which was found by \citet{Pereira2015a} under quiet-sun conditions, is no longer valid at flare ribbons.

Finally we have to emphasize that this is a 1D RHD modeling. It is unclear how the 3D radiative transfer, non-thermal electrons or magnetic fields change the synthetic Mg II line profiles at flare ribbons. Moreover, the current modeling is still incapable of reproducing Mg II line profiles with red wing enhancement. In addition, the electron density required to reproduce a single-peaked Mg II h \& k profile is still too high to reach in current modelings with an electron beam energy flux less than $10^{11}\ \mathrm{erg\ cm^{-2} s^{-1}}$, which often correspond to smaller flares with a class lower than M.

\begin{acknowledgements}

This work is supported by NSFC grants 11825301 and 11790304(11790300), the Strategic Priority Research Program of CAS with grant XDA17040507, the Research Council of Norway through its Centres of Excellence scheme, project number 262622, and through grants of computing time from the Programme for Supercomputing. H.T. acknowledges support by ISSI/ISSI-BJ to the team "Diagnosing heating mechanisms in solar flares through spectroscopic observations of flare ribbons".  IRIS is a NASA small explorer mission developed and operated by LMSAL with mission operations executed at NASA Ames Research center and major contributions to downlink communications funded by ESA and the Norwegian Space Centre. We thank Ying Li, Jie Hong and Mei Zhang for helpful discussions.

\end{acknowledgements}

\end{document}